**Cell-to-Cell stochastic fluctuations in apoptotic signaling can decide between life and death**

Subhadip Raychaudhuri[1,2,3*], Eric Willgohs[1], Thuc-Nghi Nguyen[1], Elaine M. Khan[4], Tzipora Goldkorn[3,4*]

[1] Department of Biomedical Engineering, University of California Davis
[2] Biophysics Graduate Group, University of California Davis
[3] Graduate Group in Immunology, University of California Davis
[4] Signal Transduction, School of Medicine, University of California Davis

**Apoptosis, or genetically programmed cell death, is a crucial cellular process that maintains the balance between life and death in cells. Apoptotic cell signaling, that is typically triggered by death receptor Fas (CD95) activation or stress conditions such as oxidative stress or UV radiation, is very complex due to the presence of two major pathways (type 1 and type 2) as well as numerous regulatory structures in the signaling network (1,2). The precise molecular mechanism of apoptosis signaling and how these two pathways are differentially activated under distinct apoptotic stimuli is poorly understood. We developed a Monte Carlo-based stochastic simulation model that can characterize distinct signaling behaviors in the two major pathways of apoptotic signaling using a novel probability distribution-based approach. Specifically, we show that for a weak death signal, such as low levels of death ligand Fas (CD95) binding or under stress conditions, the type 2 mitochondrial pathway dominates apoptotic signaling. Our results also show signaling in the type 2 pathway is stochastic, where the population average over many cells does not capture the cell-to-cell fluctuations in the time course (~1 – 10 hours) of downstream caspase-3 activation (3,4,5,6). On the contrary, the probability distribution of caspase-3 activation for the mitochondrial pathway shows a distinct bimodal behavior that can be used to characterize the stochastic signaling in type 2 apoptosis. Interestingly, such stochastic fluctuations in apoptosis signaling happen even in the presence of large numbers of signaling molecules (7,8). In a fluctuating environment, such stochasticity in the timecourse of caspase-3 activation may be an adaptive mechanism for allowing a competing survival signal to win over a weak death trigger before the critical cell fate decision is made and thus minimizes the risk of pathologies.**

Programmed cell death, apoptosis, is one of the most important cellular processes that is critical to a wide range of phenomena starting from normal development of multicellular organisms to maintaining homeostasis in an immune response (1,2). Any disruption of the balance in the apoptotic cell death signaling can lead to diseases ranging from cancer in the case of under-apoptosis to degenerative disorders in over-apoptosis. Two major pathways and their associated signaling molecules have been identified that mediate apoptosis cell signaling (1,2). Luo et al. have shown that two different rate constants exist

---

for caspase-3 cleavage reactions but the precise molecular mechanism behind such behavior was not clear (9). To add to this complexity, activation of the apoptosis signaling pathway, as shown by Goldkorn and co-workers (10 and references therein), also activates the competing survival signaling pathway. How does a cell decide between life and death under different apoptotic stimuli? Using a Monte Carlo-based stochastic simulation model we show that cells can use stochastic signaling through the mitochondrial type 2 pathway under weak apoptotic triggers to decide the ultimate cell fate of life or death.

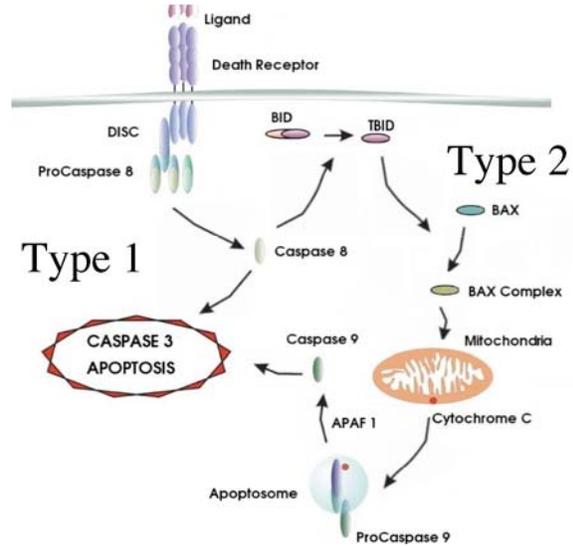

Previous attempts to mathematically model apoptosis signaling, after the pioneering work by Fussenegger et al. (3), was based on mean-field deterministic differential equations (3,4,5,6). Those previous studies provided important insights into the problem of apoptotic cell signaling. We used the kinetic parameter values used in those studies to estimate the probabilistic parameters used in our model. However, under identical cellular conditions those deterministic models would result in identical outcomes, hence there were no cell-to-cell variations in the time-course of signaling molecules and every cell would behave in the identical manner.

Figure 1. Schematics of the apoptosis signaling network

Results from our computational model show for the first time that cell-to-cell variation (stochasticity) in apoptotic signaling can be significant under certain conditions, which cannot be captured by population average (average over many cells) behavior, and may lead to the ultimate cell fate decision. Even though the investigation of such stochastic effects in gene regulatory and other types of signaling networks have attracted considerable experimental and theoretical attention, stochastic effects were thought to be important only when a low copy number of a specific molecular species (~ 1-100) in a cell is involved (7,8). On the contrary, in complex cell-signaling processes such as apoptosis, a large copy number of signaling molecules (~10,000 –100,000 in mammalian cells) is present and is thought to make the system behave in a deterministic manner. Interestingly, our results are showing that cell-to-cell fluctuations in apoptosis signaling occur even in the presence of a large number of molecules. Such stochastic effects are attributed to the complex structure of the apoptosis signaling network and low probability of formation of the apoptotsome signaling complex. We have developed a probability distribution-based (histogram plots of many single-cell observations) approach to characterize such stochastic effects in apoptotic signaling.

Our stochastic simulation belongs to the kinetic Monte Carlo class of models as we sample individual molecules randomly and choose for either diffusion or reaction move. One novel aspect of our stochastic model is that instead of using a free energy-based

Metropolis scheme we use probabilistic rate constants for reaction and diffusion of signaling molecules (11,12). The probabilistic parameters used in our model are mapped to the macroscopic experimentally measurable constants in such a manner that a detailed balance condition is satisfied at each point in space. Our stochastic computational model considers apoptotic signaling through two distinct pathways – (a) direct activation of caspase-3 by caspase-8 (type 1) and (b) activation of caspase-3 by mitochondrial cytochrome-c release and apoptosome formation (type 2) (Figure 1). Intracellular apoptosis signaling in our model is triggered by the activation of caspase-8 molecules at the cell surface, which in turn diffuse in the cytosol and activate both pathways of apoptosis signaling (Figure 1). In the type 1 pathway, caspase-8 molecules directly catalyze the cleavage reaction of procaspase-3 to generate caspase-3. In the type 2 pathway, caspase-8 binds with Bid and catalyzes its truncation to form tBid, which in turn binds to Bax to generate Bax2 complex molecules. Bcl2, an anti-apoptotic molecule, can inhibit both tBid and Bax, thus creating a local loop structure in the type 2 signaling cascade. The Bax2 complex leads to cytochrome c release from mitochondria and ultimately to the activation of caspase-3 (Figure 1). Hence, caspase-3 activation at the end of both the type 1 and the type 2 pathways creates a global loop structure in apoptosis signaling. In order to investigate cell death under varying strengths of apoptotic stimuli, we changed the concentrations of procaspase-8 molecules and explored its effect on the downstream signaling for the two different pathways. Caspase-3 cleavage is irreversible and leads to apoptosis. Caspase-3 activation is considered the readout for apoptosis signaling that decides the final cell fate. We followed the dynamics of signaling reactions by measuring the number of signaling molecules at regular time intervals. Each run of our simulation corresponds to apoptosis signaling observed at the single cell level.

Activation of downstream signaling molecules in type 1 and type 2 pathways and differential activation of those two pathways under differential cellular conditions were previously not well understood (9). By setting the appropriate kinetic constants to zero we are also able to activate only one of the two pathways. For the type 1 pathway, the population average over many cells (average over many runs in our simulations) can capture the essential dynamics of direct caspase8-caspase3 activation (Figure 2A). Results of a mean-field model, based on kinetic rate equations of signaling reactions (solved using an ODE solver in the MATLAB software), are also consistent with the observed average behavior of stochastic simulations. Probability distribution of activated caspase-3 for various time points show a gradual increase (Figure 3A). In stark contrast to the type 1 pathway, for type 2 the Caspase-3 activation in individual cell simulations varied by an order of magnitude from minutes to hours (Figure 2B). Clearly, few cells with very slow activation of caspase-3 can shift the average time for peak caspase-3 activation to ~10 hours. Such slow caspase-3 activation and subsequent apoptosis have been observed in experiments where apoptotic signaling was triggered by stress conditions such as an oxidative agent (10) and must be dominated by the type 2 pathway of apoptosis.

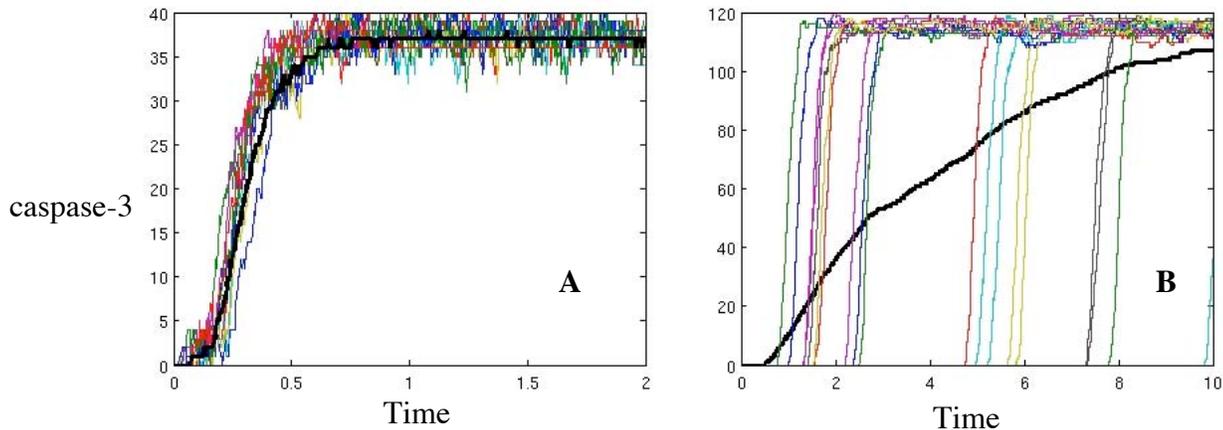

Figure 2. Time-course of caspase-3 activation is shown for - (A) Type 1 pathway (B) Type 2 pathway. The solid black curves represent average over many (~100) cells. Different colors correspond representative individual cells. (Unit time step = 100 seconds).

Another striking feature of caspase-3 activation for the type 2 pathway is that the completion of caspase-3 activation, once initiated, is relatively quick (~minutes) compared to the timescale over which cell-to-cell variations occur (~hours). Interestingly, the average for numerous simulations does not capture the cell-to-cell fluctuations that are observed in individual stochastic simulations (Figure 2B). The ratio, variance/average (also called the fano factor), for the number of activated caspase-3 molecules stays large (> 1) for a long time (~ hours). Stochastic fluctuation in the time course of caspase-3 activation initiation coupled with the rapid activation of caspase-3 once it is initiated, leads to a characteristic bimodal behavior for the probability distribution of caspase-3 activation (Figure 3B). One characteristic feature of this bimodal probability distribution is that it is not symmetric under time reversal (t → -t), which is a signature of a truly non-equilibrium process.

To investigate the cause of fluctuations in caspase-3 activation for the type 2 pathway we considered the activation of downstream signaling molecules in that pathway. Our simulations show small fluctuations in Bax activation and subsequent Cytochrome-c release from the mitochondria lead to large (order of magnitude increase) fluctuations in apoptosome formation. Such cell-to-cell stochastic fluctuation in the time-course of apoptosome formation mainly drives the fluctuations in caspase-9 and hence caspase-3 activation because rapid activation of those molecules is observed immediately after the formation of apoptosomes. In our model, an apoptosome is formed only when two Apaf molecules are bound to a Cytochrome-c molecule and caspase-9 is activated only through binding with an intact apoptosome. Apoptosome formation is a low probability event due to the low binding on-rate of cytochrome c and Apaf and the random diffusion of a three-molecule complex into the same spatial location. Interestingly, only 1-5 molecules of apoptosome are sufficient to activate the downstream caspase molecules caspase-9 and caspase-3. Hence for this type 2 pathway, the final caspase-3 activation can be modeled by a stochastic differential equation of the form

$$\frac{d[caspase3]}{dt} \approx -(V_2 + \eta(t))[caspase3],$$

where the noise term η carries the information of a time-dependent noise in a stochastic rate constant. Such a stochastic rate-constant can arise, either due to competitive binding of a specific signaling molecule such as Bcl2, or low probability of a specific reaction as seen in apoptosome formation, in the type 2 pathway. The above equation clearly shows that even in the presence of a large number of caspase-3 molecules, stochastic effects persist.

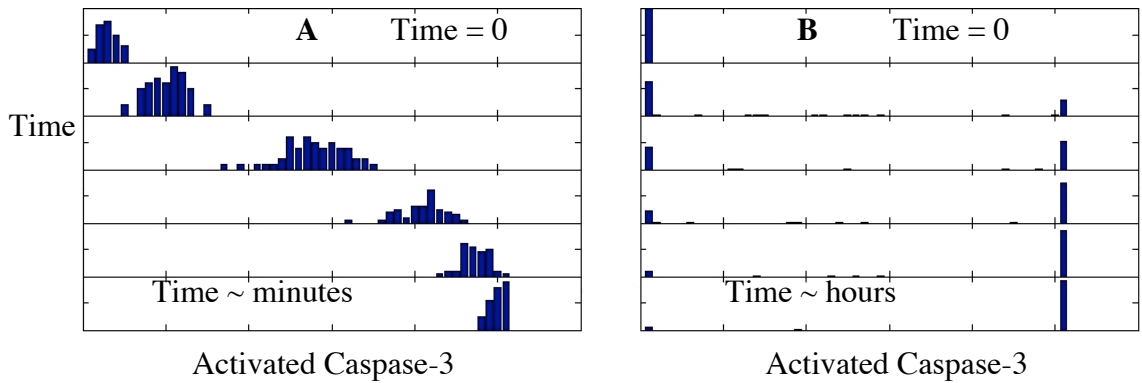

Figure 3. Distribution of activated caspase-3 molecules for various time points. (a) gradual increase in caspase-3 for the type 1 pathway (b) bimodal distribution for the type 2 pathway.

When both type 1 and type 2 pathways are combined, the presence of a large number of procaspase-8 (~ 100 nanomolar or more) made the type 1 pathway dominant by rapid activation of caspase-3. However, as the concentration of procaspase-8 is lowered, the type 2 pathway dominates because few caspase-8 molecules preferentially bind to Bid (larger on rate) compared to caspase-3 molecules. Note, the details of reaction kinetics does not change our main conclusions as long as this strong preferential binding to Bid holds true and the network structure of the apoptotic signaling pathway remains the same (Figure 1). Thus the number of caspase-8 molecules, or equivalently the strength of the apoptotic stimulus, can be used as a control to selectively activate the type1 or type 2 pathway. When the number of caspase-8 molecules is less than one thousand, mainly the type 2 pathway is activated and characteristic type 2 fluctuations, as shown in Figure 2b and Figure 3b, are again observed. This characteristic bimodal nature of the caspase-3 probability distribution can be used to detect contributions from the type 2 pathway. Our simulations also elucidate that the variation in Bcl2 levels in the type 2 pathway leads to nonlinear changes in caspase-3 activation. Competitive binding of Bcl2 to tBid and Bax creates a local loop structure in the type2 pathway that controls Bax2 complex formation and subsequent nonlinear stochastic fluctuations in caspase-3 activation.

Apoptosis signaling is one of the most investigated cell signaling pathways because of its implications for various diseases and also as a target for chemotherapeutic interventions (1,13). Major signaling molecules of the apoptotic pathway as well as the network structure of the signaling reactions are known from biological experiments. Mathematical modeling of apoptosis signaling has so far considered a set of kinetic rate equations of signaling reactions and nonlinear effects of specific signaling molecules were analyzed (3,4). All-or-none type bistable behavior in apoptosis cell signaling experiments (14) was explained by fluctuations in environmental conditions (5) and low probability of apoptosome formation (6). In contrast to these earlier studies, our stochastic simulations of apoptosis clearly show that a cell fate decision can vary even under identical cellular and environmental conditions and a probability distribution-based approach is needed to characterize such stochastic effects in cellular signaling. Specifically, our results show that, for a weak apoptotic stimulus, cells use the mitochondrial type 2 pathway and cell-to-cell fluctuations in the caspase-3 activation can vary 1 hour - 10 hours even under identical cellular and environmental conditions. In the presence of a competing survival signal, which can happen under oxidative stress induced apoptosis (10), the stochastic rate constant term in equation (1) for the type 2 pathway is modified and can result in cell survival. Such a strategy may be used by nature to adjust to a fluctuating environment where few cells wait a long time before activating downstream apoptosis signaling so that any competing survival signal gets a chance to win over the weak apoptotic stimulus and thus maintains a balance between life and death.

**Acknowledgement**
We thank Matthew Lauer for helping with the programming and creating graphs. We thank Anthony Tam for critical reading of the manuscript and comments.